\documentclass[journal=jacsat,manuscript=letter]{achemso}

\usepackage[version=3]{mhchem} 
\usepackage{comment}

\author{Thomas Zacharias}
\altaffiliation{T.Z. and R.G. contributed equally to this work. }
\author{Robert Gray}
\altaffiliation{T.Z. and R.G. contributed equally to this work. }
\author{Ryoto Sekine}
\author{James Williams}
\author{Selina Zhou}
\author{Alireza Marandi}
\email{marandi@caltech.edu}
\affiliation[California Institute of Technology]
{Department of Electrical Engineering, California Institute of Technology, Pasadena, California, 91125, USA.}
\alsoaffiliation[California Institute of Technology]
{Department of Applied Physics, California Institute of Technology, Pasadena, California, 91125, USA.}

\title[]
  {Energy-Efficient Ultrashort-Pulse Characterization using Nanophotonic Parametric Amplification}

\abbreviations{Ultrafast optics, Integrated photonics, Ultrashort pulse characterization}

\begin{document}

\begin{abstract}
The growth of ultrafast nanophotonic circuits necessitates the development of energy-efficient on-chip pulse characterization techniques. 
Nanophotonic realizations of Frequency Resolved Optical Gating (FROG), a common pulse characterization technique in bulk optics, have been challenging due to their non-collinear nature and the lack of efficient nonlinear optical processes in the integrated platform. 
Here, we experimentally demonstrate a novel FROG-based technique compatible with the nanophotonic platform that leverages the high gain-bandwidth of a dispersion-engineered degenerate optical parametric amplifier (DOPA) for energy-efficient ultrashort pulse characterization. 
We demonstrate on-chip pulse characterization of sub-80-fs, {\raise.17ex\hbox{$\scriptstyle\sim$}} 1-fJ pulses using just {\raise.17ex\hbox{$\scriptstyle\sim$}} 60-fJ of gate pulse energy, which is several orders of magnitude lower than the gate pulse energy required for characterizing similar pulses in the bulk counterpart.
In the future, we anticipate our work will enable the characterization of ultraweak-ultrashort pulses with energies at the single photon level. 
\end{abstract}

\section*{Introduction}

Ultrafast integrated photonics aims to bring the advantages of ultrashort pulses largely limited to expensive, bulky optical systems to scalable, compact nanophotonic platforms \cite{chang2022integrated}. 
Developments in this field have so far been focused on either generating ultrafast pulses on-chip \cite{guo2023ultrafast, yu2022integrated, davenport2018integrated, yuan2023soliton} or on leveraging ultrashort pulses for applications in time-keeping \cite{newman2019architecture}, quantum information processing \cite{nehra2022few}, and computing \cite{guo2022femtojoule}. 
Demonstrations of ultrashort pulse characterization techniques on nanophotonic platforms \cite{yu2023frequency, pasquazi2011sub} have been limited due to the requirements of strong optical nonlinearity and the non-collinear nature of many typical pulse characterization techniques. 
Ultrashort pulse characterization is a critical tool for leveraging the unique properties of ultrashort pulses - short pulse width, high peak power, and high repetition rate - for probing ultrafast phenomena \cite{riek2017subcycle}, enhancing nonlinear interactions \cite{wegener2006extreme}, and increasing information density for ultrafast information processing \cite{li2024deep}. 
Developing ultrashort pulse characterization techniques in nanophotonics is an important step toward developing integrated ultrafast photonic systems. 
Additionally, the challenges associated with off-chip temporal characterization, weak pulse energies, and temporal distortions resulting from pulse extraction necessitate the need for energy-efficient ultrashort on-chip pulse characterization techniques. 

Here, we experimentally demonstrate a novel pulse characterization technique compatible with integrated photonics for energy-efficient on-chip pulse characterization. 
Our technique uses dispersion-engineered optical parametric amplifiers (OPAs) in lithium niobate nanophotonic waveguides combined with a frequency-resolved optical gating (FROG) - based retrieval algorithm \cite{trebino2000frequency} for on-chip pulse characterization.  
Cross-FROG (XFROG) based on non-collinear OPAs has been previously demonstrated in bulk optics as one of the most sensitive ultrashort pulse characterization techniques \cite{zhang2003measurement}. 
However, its non-collinear nature makes it incompatible with nanophotonics.
Additionally, fJ pulse characterization using the bulk OPA-XFROG has required tens of $\mu$J of gate pulse energies \cite{zhang2003measurement, zhang2004sub}. 
Here, we overcome these limitations by operating in the degenerate and collinear regime, making our technique compatible with nanophotonic realization. 
Moreover, on-chip degenerate OPAs provide unparalleled gain, especially in nanophotonic lithium niobate with low-energy pump pulses \cite{ledezma2022intense}. 
A combination of the spatial mode confinement due to the waveguides and temporal mode confinement due to dispersion engineering enables gain-bandwidth levels unavailable to bulk crystals.
This results in operating the Degenerate Optical Parametric Amplifier (DOPA) - XFROG using a fraction of the pump energy compared to the bulk counterpart. 
Our scheme paves the way toward the measurement of ultrashort-ultraweak optical pulses that were not possible with bulk crystals. 

\section*{Methods}
\subsection*{Principle of operation for DOPA-XFROG}
The FROG \cite{trebino2000frequency} uses an iterative 2D phase-retrieval algorithm to recover the intensity and phase of an unknown optical field from an intensity spectrogram. 
The phase retrieval uses a generalized projections optimization algorithm that iteratively searches for the electric field that can create the intensity spectrogram while satisfying an optical nonlinearity constraint.
In our DOPA-XFROG, the optical nonlinearity is defined by the degenerate optical parametric amplification process which can be mathematically modeled as 

\begin{equation}
    E^{DOPA}(t, \tau) = E(t)\cosh(\kappa |G(t-\tau)|) + iE^*(t)\sinh(\kappa|G(t-\tau)|)\exp(i\angle G(t-\tau)),        
    \label{Eq: FieldDOPA}
\end{equation}

\noindent where $E^{DOPA}(t, \tau)$ is the pulse field resulting from the nonlinear process, $E(t)$ is the unknown signal pulse field to be measured, $G(t)$ is the known gate pulse field that pumps the OPA, and $\tau$ is the time delay between the signal and the pump. 
$\kappa^2 = \frac{2\omega_s^2 d_{eff}^2z^2}{n_s^2 n_p\epsilon_0 c^3 A_{eff}}$ is the gain parameter for the OPA where $d_{eff} = \frac{2}{\pi}  d_{33}$ is the effective nonlinear coefficient \cite{ledezma2022intense}, z is the length of the nonlinear interaction process, and $n_s, n_p$ is the effective index of the signal and pump respectively. 
The carrier frequency of the gate pulse is twice that of the unknown pulse. 
The degenerate OPA process is sensitive to the relative phase of the signal and gate pulses which determines whether the signal to be measured is amplified through parametric gain or deamplified through second harmonic generation. 
The dependence of the OPA process on the relative signal and gate phases within the order of an optical cycle makes the output field sensitive to experimental fluctuations. 
These include non-idealities such as stage nonrepeatability and nonlinearity, timing jitter, temperature fluctuations, alignment fluctuations, etc.
These fluctuations are calibrated using a modified spectrogram measurement technique that allows for the simultaneous collection of a calibration signal. 
The spectrogram is created by continuously scanning the stage while detecting the output signal one frequency component at a time on an optical spectrum analyzer. 
The output signal is simultaneously collected in a slow detector that acts as the calibration signal. 
The slow detector is expected to see the same signal for each consecutive scan of the delay stage and can therefore be used for calibration. 
The calibrated spectrogram can then be passed through the custom DOPA-XFROG algorithm for pulse retrieval.

The custom recovery algorithm uses generalized projections algorithm \cite{trebino2000frequency} to enforce the mathematical constraint defined in Eq. \ref{Eq: FieldDOPA} through iterative gradient descent. 
The corresponding gradients were analytically derived from Eq. \ref{Eq: FieldDOPA} and can be found in the supplementary information. 
Equation \ref{Eq: FieldDOPA} accurately models the amplification process when operating in the quasistatic regime - where dispersion and walk-off are negligible. 
Dispersion-engineering, unavailable in bulk crystals, achieved through precise control over waveguide geometry enables operating nanophotonic OPAs in the quasistatic regime \cite{ledezma2022intense, jankowski2022quasi} thus ensuring the validity of Eq. \ref{Eq: FieldDOPA}. 
\begin{figure}[h] 
	\begin{centering}
    		\includegraphics[width=1\linewidth,trim={16.7cm 16.4cm 21.4cm 5.9cm},clip]{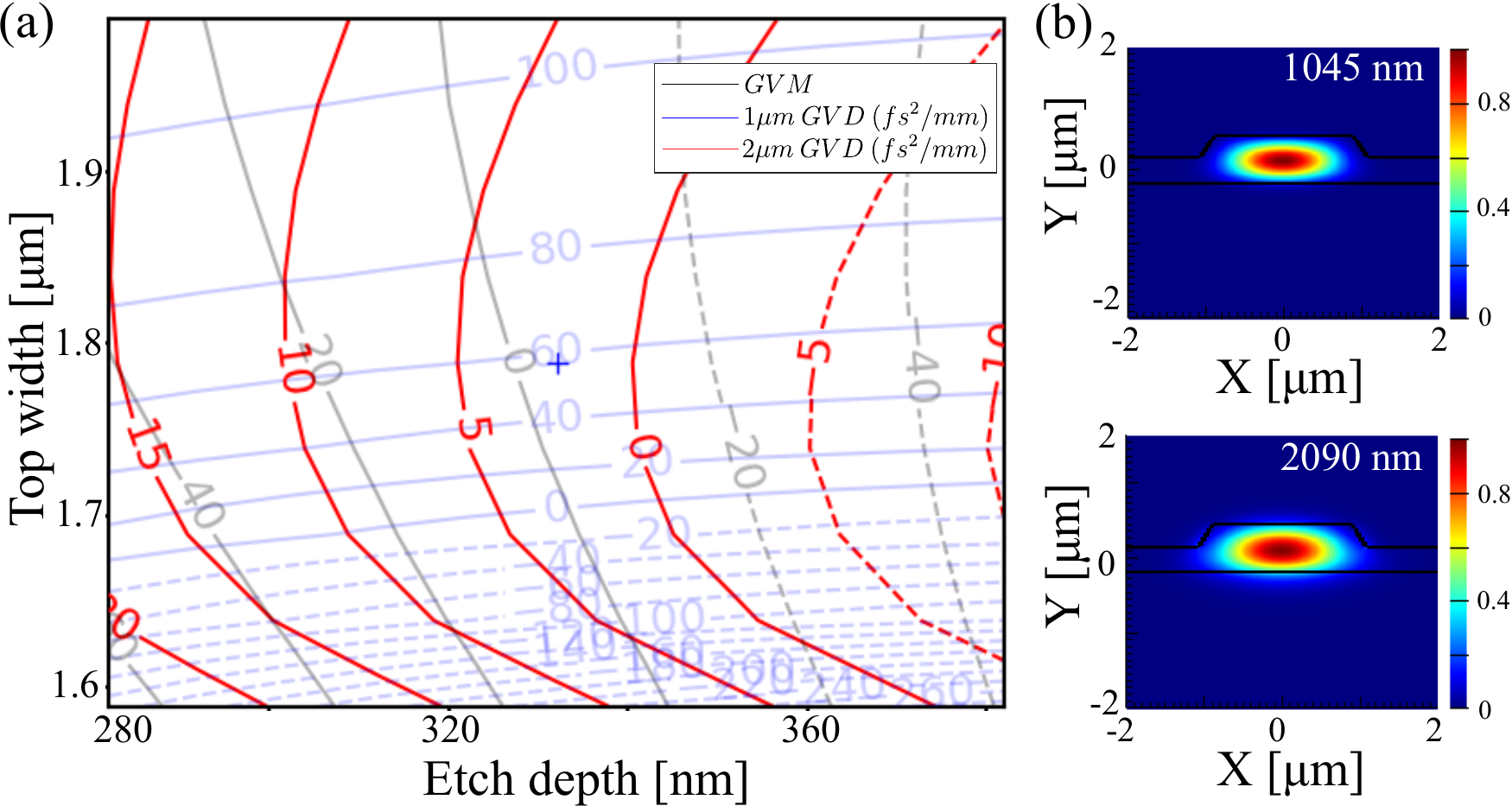}
    	\par\end{centering}
        \caption{ \textbf{(a) Dispersion map.} Simulated GVM between 1045 nm and 2090 nm, GVD at 1045 nm, and GVD at 2090 nm while scanning for different top widths and etch depths. The blue cross corresponds to the dimensions of the etched waveguide. \textbf{(b) Mode simulations.} Electric field profiles of the fundamental TE modes at pump and signal wavelengths.}
	\label{fig:dispMap}
\end{figure}

\begin{figure*}[t] 
	\begin{centering}
    		\includegraphics[width=1\linewidth,trim={0 0 0 0},clip]{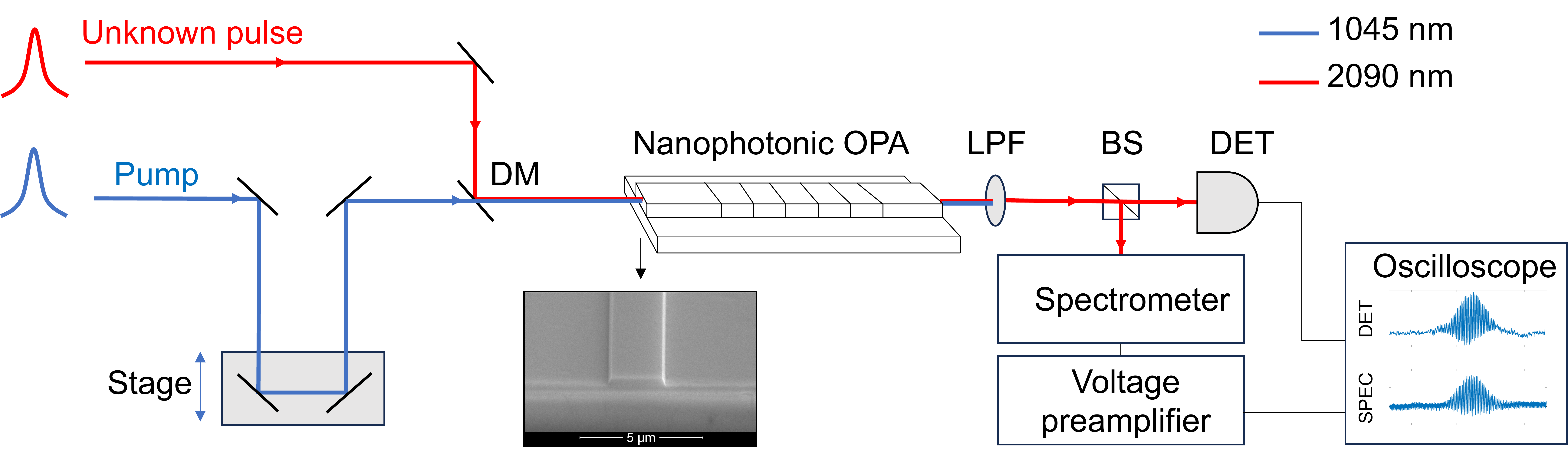}
    	\par\end{centering}

        \caption{ \textbf{Experimental setup.} DM: Dichroic Mirror; OPA: Optical Parametric Amplifier; LPF: Low Pass Filter; BS: Beam Splitter; DET: Detector. Inset: Scanning electron microscope image of chip facet.}
	\label{fig:sup-workflow}
\end{figure*}

\begin{figure*}[t] 
	\begin{centering}
    		\includegraphics[width=1\linewidth,trim={2cm 6.5cm 6.5cm 2cm},clip]{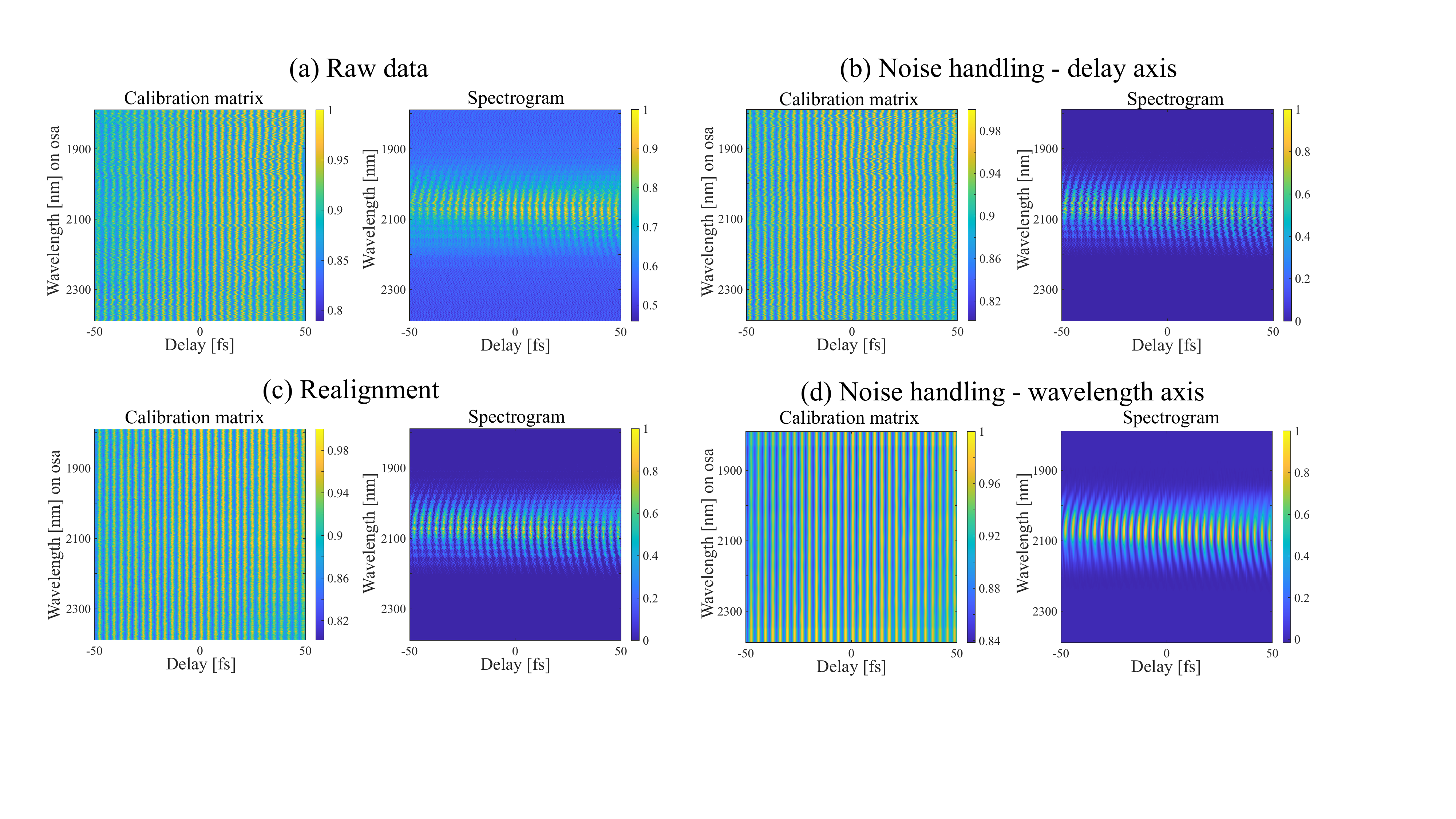}
    	\par\end{centering}
        \caption{ \textbf{Postprocessing.} Each sub-panel displays the calibration matrix created from the detector signal and the spectrogram created from the spectrometer signal at each step of the postprocessing.
        (a) displays the raw data collected from the experimental measurement. 
        (b) displays the matrices after lowpass filtering along the
        delay axes for both the matrices and thresholding the spectrogram.
        (c) displays the result of realignment and linearization. 
        (d) displays the result of lowpass filtering the wavelength axis and renormalization. 
    	}
	\label{fig:sup-postproc}
\end{figure*}
\begin{figure*}[t] 
	\begin{centering}
    		\includegraphics[width=1\linewidth,trim={3cm 0.5cm 3cm 1cm},clip]{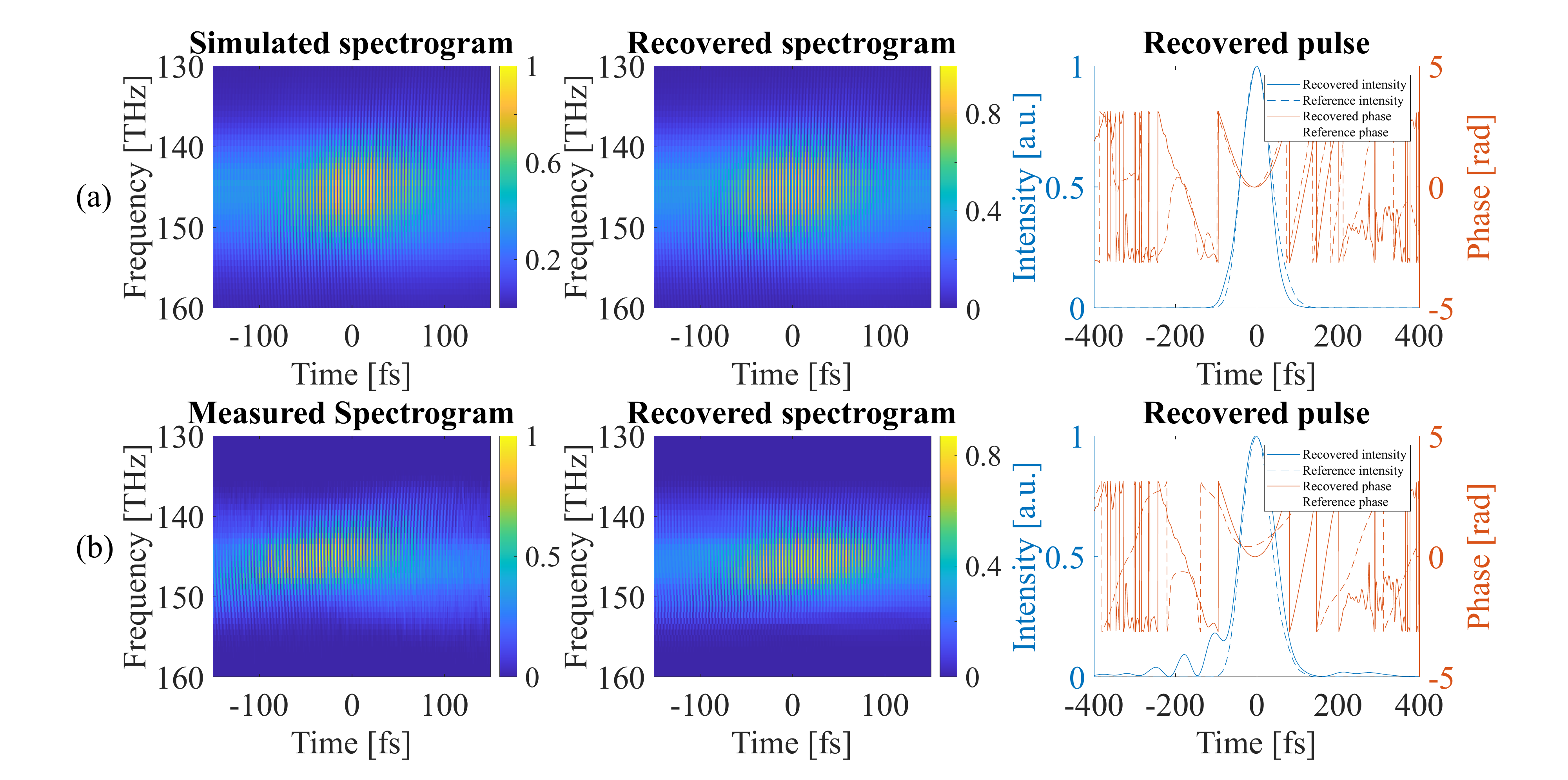}
    	\par\end{centering}
        \caption{ \textbf{Result.} \textbf{(a) Simulation } (Left) Simulated spectrogram for a reference pulse. (Center) Recovered spectrogram using DOPA-XFROG algorithm. (Right) Comparison of recovered pulse profile with reference pulse. 
        \textbf{(b) Experiment.} (Left) Experimentally measured spectrogram. (Center) Recovered spectrogram using DOPA-XFROG algorithm. (Right) Comparison of recovered pulse profile with expected profile (measured with a tabletop FROG). 
    	}
	\label{fig:Results}
\end{figure*} 

\subsection*{Device design and fabrication}
The design and fabrication of the dispersion-engineered nanophotonic OPA used here follow the techniques described in \cite{ledezma2022intense}.
The waveguides were designed to have minimal dispersion and walk-off between the signal at 2090 nm and the pump at 1045 nm.   
The waveguides were fabricated on a 704-nm thin film of lithium niobate on a silica substrate and were measured using atomic force microscopy to have a top width of 1790 nm and an etch depth of 330 nm. 
The etched waveguide geometry was simulated using Lumerical to estimate a group velocity mismatch between 1045 nm and 2090 nm of 1.7 fs/mm, group velocity dispersion around 2090 nm of -3 fs$^2$/mm, group velocity dispersion around 1045 nm of 54 fs$^2$ /mm (see Fig. \ref{fig:dispMap}). 
The total length of the waveguide was 10 mm with a 4-mm-long periodically poled region and a 5.2 $\mu$m poling period.
The OPA was measured to have a gain of 188 dB/cm for a pump pulse energy of {\raise.17ex\hbox{$\scriptstyle\sim$}} 4 pJ and bandwidth of 40.5 THz using the techniques described in \cite{ledezma2022intense}.

\subsection*{Experimental Scheme}
Figure \ref{fig:sup-workflow} shows the experimental setup used for DOPA-XFROG. 
A 100-fs mode-locked fiber laser centered at 1045 nm pumps a home-built optical parametric oscillator (OPO) and the DOPA-XFROG. 
The output of the OPO is at 2090 nm which is used as the "unknown pulse" to be characterized. 
Both the pump and OPO output are characterized with a traditional table-top XFROG to create the reference pulses seen in Fig. \ref{fig:Results} (right). 
The unknown pulse is gated by the pump in the nanophotonic OPA for different delays controlled by a linearized motor stage. 
The resulting signal from the OPA is low-pass filtered and collected in a spectrometer and slow detector. 
The stage is set to scan continuously and the spectrometer measures the spectrum as a function of delay one wavelength component at a time from 1790 nm to 2390 nm with a 2 nm resolution.   
The analog output of the spectrometer is amplified using a voltage preamplifier and sent to an oscilloscope. 
The signal collected by the slow detector used for calibration during postprocessing is also sent to the same oscilloscope. 
The oscilloscope is set to trigger data collection when the slow detector signal passes a threshold voltage level indicating temporal overlap of the two pulses for each scan of the delay stage.
The spectrometer updates the collected wavelength center between subsequent scans by increments of 2 nm. 
The detector and spectrometer traces are saved on the oscilloscope for each stage scan to create the calibration matrix and raw spectrogram.
Fig. \ref{fig:sup-postproc} illustrates the postprocessing algorithm used for calibrating the spectrogram. 
The calibration process includes noise handling and realignment. 
The noise handling accounts for electronic noise and optical power fluctuations and is achieved through low-pass filtering, thresholding, and renormalization.
The realignment procedure accounts for optical phase variations caused by timing jitter between pulses, stage non-repeatability/nonlinearity, etc. 
Additional details about the postprocessing can be found in the supplementary information. 
The calibrated spectrogram (Fig. \ref{fig:sup-postproc} d, right) can now be used with the DOPA-XFROG recovery algorithm to recover the intensity and phase of the unknown pulse. 

\section*{Results}
The custom algorithm was tested by simulating the DOPA-XFROG. 
Fig. \ref{fig:Results} (a, left) shows a spectrogram simulated using the experimental pulse profiles (measured using a standard tabletop XFROG) and the nonlinear equation defined by Eq. \ref{Eq: FieldDOPA}.
This spectrogram was passed through the custom recovery algorithm which converged to the recovered spectrogram in Fig. \ref{fig:Results} (a, center) with an RMS error of $0.00048$ after $250$ iterations.
The intensity and phase of the recovered field along with the expected profile for comparison are displayed in Fig. \ref{fig:Results} (a, right). 
The phase-sensitive nature of the process can be seen in the intensity spectrogram in the form of periodic fringes. 
The frequency dependence of the fringe locations marked by the curvature along the frequency axes in the intensity spectrogram is a sign of using chirped pulses. 

Fig. \ref{fig:Results} (b, left), shows the experimentally measured spectrogram after postprocessing using a {\raise.17ex\hbox{$\scriptstyle\sim$}} 1-fJ pulse at 2090 nm gated with {\raise.17ex\hbox{$\scriptstyle\sim$}} 60-fJ pulse at 1045 nm. 
The pulse energies were estimated on-chip using the previously demonstrated input coupling characterization technique \cite{ledezma2022intense}. 
The spectrogram was passed to the DOPA-XFROG algorithm which recovered the spectrogram displayed in Fig. \ref{fig:Results} (b, center) after $500$ iterations with an error of $0.003$. 
Fig. \ref{fig:Results} (b, right), shows a comparison of the recovered pulse intensity and phase with the expected pulse profile. 
Fig. \ref{fig:Results} demonstrates that the DOPA-XFROG can successfully characterize ultrashort pulses. 
Differences between the recovered and reference pulse intensity and phase profiles can be attributed to non-idealities such as waveguide dispersion, walk-off, phase-mismatch, parasitic nonlinear processes, etc.
\section*{Discussion and Conclusion}
We developed a novel pulse characterization technique, named DOPA-XFROG, and demonstrated how it can be used for ultrafast pulse measurements in a nanophotonic lithium niobate chip. 
The spatio-temporal confinement of dispersion-engineered nanophotonics enables leveraging the high gain-bandwidth OPAs for energy-efficient pulse characterization using weak gate pulses. 
Measurements of fJ pulses in bulk OPA-XFROG setups typically required tens of $\mu$J of gate pulse energies \cite{zhang2003measurement, zhang2004sub}. 
In comparison, we demonstrate pulse characterization of {\raise.17ex\hbox{$\scriptstyle\sim$}} 1-fJ pulses using gate pulses with energy as low as {\raise.17ex\hbox{$\scriptstyle\sim$}} 60-fJ.

Attempts at experimentally measuring weaker pulses are currently limited by parasitic nonlinear effects due to gain saturation that can lead to spectral broadening and deviation of the parametric amplification process from the mathematical model defined in Eq. \ref{Eq: FieldDOPA}.
Using shorter OPAs with larger pump pulse energies may enable the measurement of sub-atto-Joule pulses on-chip.
This technique could be especially useful for system integration where a small amount of pulse energy could be tapped from a highly integrated photonic circuit for further on-chip pulse characterization.

\begin{acknowledgement}

Device nanofabrication was performed at the Kavli Nanoscience Institute (KNI) at Caltech. The authors gratefully acknowledge support from ARO grant no. W911NF-23-1-0048, NSF grant no. 1918549, AFOSR award FA9550-23-1-0755, DARPA award D23AP00158, the Center for Sensing to Intelligence at Caltech, the Alfred P. Sloan Foundation, and NASA/JPL.

\end{acknowledgement}

\begin{suppinfo}
Additional details about the experimental scheme, postprocessing, and recovery algorithm are provided in the supplementary information.
\end{suppinfo}

\bibliography{references}

\providecommand{\latin}[1]{#1}
\makeatletter
\providecommand{\doi}
  {\begingroup\let\do\@makeother\dospecials
  \catcode`\{=1 \catcode`\}=2 \doi@aux}
\providecommand{\doi@aux}[1]{\endgroup\texttt{#1}}
\makeatother
\providecommand*\mcitethebibliography{\thebibliography}
\csname @ifundefined\endcsname{endmcitethebibliography}  {\let\endmcitethebibliography\endthebibliography}{}
\begin{mcitethebibliography}{19}
\providecommand*\natexlab[1]{#1}
\providecommand*\mciteSetBstSublistMode[1]{}
\providecommand*\mciteSetBstMaxWidthForm[2]{}
\providecommand*\mciteBstWouldAddEndPuncttrue
  {\def\EndOfBibitem{\unskip.}}
\providecommand*\mciteBstWouldAddEndPunctfalse
  {\let\EndOfBibitem\relax}
\providecommand*\mciteSetBstMidEndSepPunct[3]{}
\providecommand*\mciteSetBstSublistLabelBeginEnd[3]{}
\providecommand*\EndOfBibitem{}
\mciteSetBstSublistMode{f}
\mciteSetBstMaxWidthForm{subitem}{(\alph{mcitesubitemcount})}
\mciteSetBstSublistLabelBeginEnd
  {\mcitemaxwidthsubitemform\space}
  {\relax}
  {\relax}

\bibitem[Chang \latin{et~al.}(2022)Chang, Liu, and Bowers]{chang2022integrated}
Chang,~L.; Liu,~S.; Bowers,~J.~E. Integrated optical frequency comb technologies. \emph{Nature Photonics} \textbf{2022}, \emph{16}, 95--108\relax
\mciteBstWouldAddEndPuncttrue
\mciteSetBstMidEndSepPunct{\mcitedefaultmidpunct}
{\mcitedefaultendpunct}{\mcitedefaultseppunct}\relax
\EndOfBibitem
\bibitem[Guo \latin{et~al.}(2023)Guo, Gutierrez, Sekine, Gray, Williams, Ledezma, Costa, Roy, Zhou, Liu, \latin{et~al.} others]{guo2023ultrafast}
Guo,~Q.; Gutierrez,~B.~K.; Sekine,~R.; Gray,~R.~M.; Williams,~J.~A.; Ledezma,~L.; Costa,~L.; Roy,~A.; Zhou,~S.; Liu,~M.; others Ultrafast mode-locked laser in nanophotonic lithium niobate. \emph{Science} \textbf{2023}, \emph{382}, 708--713\relax
\mciteBstWouldAddEndPuncttrue
\mciteSetBstMidEndSepPunct{\mcitedefaultmidpunct}
{\mcitedefaultendpunct}{\mcitedefaultseppunct}\relax
\EndOfBibitem
\bibitem[Yu \latin{et~al.}(2022)Yu, Barton~III, Cheng, Reimer, Kharel, He, Shao, Zhu, Hu, Grant, \latin{et~al.} others]{yu2022integrated}
Yu,~M.; Barton~III,~D.; Cheng,~R.; Reimer,~C.; Kharel,~P.; He,~L.; Shao,~L.; Zhu,~D.; Hu,~Y.; Grant,~H.~R.; others Integrated femtosecond pulse generator on thin-film lithium niobate. \emph{Nature} \textbf{2022}, \emph{612}, 252--258\relax
\mciteBstWouldAddEndPuncttrue
\mciteSetBstMidEndSepPunct{\mcitedefaultmidpunct}
{\mcitedefaultendpunct}{\mcitedefaultseppunct}\relax
\EndOfBibitem
\bibitem[Davenport \latin{et~al.}(2018)Davenport, Liu, and Bowers]{davenport2018integrated}
Davenport,~M.~L.; Liu,~S.; Bowers,~J.~E. Integrated heterogeneous silicon/III--V mode-locked lasers. \emph{Photonics Research} \textbf{2018}, \emph{6}, 468--478\relax
\mciteBstWouldAddEndPuncttrue
\mciteSetBstMidEndSepPunct{\mcitedefaultmidpunct}
{\mcitedefaultendpunct}{\mcitedefaultseppunct}\relax
\EndOfBibitem
\bibitem[Yuan \latin{et~al.}(2023)Yuan, Gao, Yu, Wang, Jin, Ji, Feshali, Paniccia, Bowers, and Vahala]{yuan2023soliton}
Yuan,~Z.; Gao,~M.; Yu,~Y.; Wang,~H.; Jin,~W.; Ji,~Q.-X.; Feshali,~A.; Paniccia,~M.; Bowers,~J.; Vahala,~K. Soliton pulse pairs at multiple colours in normal dispersion microresonators. \emph{Nature Photonics} \textbf{2023}, \emph{17}, 977--983\relax
\mciteBstWouldAddEndPuncttrue
\mciteSetBstMidEndSepPunct{\mcitedefaultmidpunct}
{\mcitedefaultendpunct}{\mcitedefaultseppunct}\relax
\EndOfBibitem
\bibitem[Newman \latin{et~al.}(2019)Newman, Maurice, Drake, Stone, Briles, Spencer, Fredrick, Li, Westly, Ilic, \latin{et~al.} others]{newman2019architecture}
Newman,~Z.~L.; Maurice,~V.; Drake,~T.; Stone,~J.~R.; Briles,~T.~C.; Spencer,~D.~T.; Fredrick,~C.; Li,~Q.; Westly,~D.; Ilic,~B.~R.; others Architecture for the photonic integration of an optical atomic clock. \emph{Optica} \textbf{2019}, \emph{6}, 680--685\relax
\mciteBstWouldAddEndPuncttrue
\mciteSetBstMidEndSepPunct{\mcitedefaultmidpunct}
{\mcitedefaultendpunct}{\mcitedefaultseppunct}\relax
\EndOfBibitem
\bibitem[Nehra \latin{et~al.}(2022)Nehra, Sekine, Ledezma, Guo, Gray, Roy, and Marandi]{nehra2022few}
Nehra,~R.; Sekine,~R.; Ledezma,~L.; Guo,~Q.; Gray,~R.~M.; Roy,~A.; Marandi,~A. Few-cycle vacuum squeezing in nanophotonics. \emph{Science} \textbf{2022}, \emph{377}, 1333--1337\relax
\mciteBstWouldAddEndPuncttrue
\mciteSetBstMidEndSepPunct{\mcitedefaultmidpunct}
{\mcitedefaultendpunct}{\mcitedefaultseppunct}\relax
\EndOfBibitem
\bibitem[Guo \latin{et~al.}(2022)Guo, Sekine, Ledezma, Nehra, Dean, Roy, Gray, Jahani, and Marandi]{guo2022femtojoule}
Guo,~Q.; Sekine,~R.; Ledezma,~L.; Nehra,~R.; Dean,~D.~J.; Roy,~A.; Gray,~R.~M.; Jahani,~S.; Marandi,~A. Femtojoule femtosecond all-optical switching in lithium niobate nanophotonics. \emph{Nature Photonics} \textbf{2022}, \emph{16}, 625--631\relax
\mciteBstWouldAddEndPuncttrue
\mciteSetBstMidEndSepPunct{\mcitedefaultmidpunct}
{\mcitedefaultendpunct}{\mcitedefaultseppunct}\relax
\EndOfBibitem
\bibitem[Yu \latin{et~al.}(2023)Yu, Lun, Lin, Li, Huang, Liu, Wu, Wang, Cheng, Li, \latin{et~al.} others]{yu2023frequency}
Yu,~H.; Lun,~Y.; Lin,~J.; Li,~Y.; Huang,~X.; Liu,~B.; Wu,~W.; Wang,~C.; Cheng,~Y.; Li,~Z.-y.; others Frequency-Resolved Optical Gating in Transverse Geometry for On-Chip Optical Pulse Diagnostics. \emph{Laser \& Photonics Reviews} \textbf{2023}, \emph{17}, 2201017\relax
\mciteBstWouldAddEndPuncttrue
\mciteSetBstMidEndSepPunct{\mcitedefaultmidpunct}
{\mcitedefaultendpunct}{\mcitedefaultseppunct}\relax
\EndOfBibitem
\bibitem[Pasquazi \latin{et~al.}(2011)Pasquazi, Peccianti, Park, Little, Chu, Morandotti, Aza{\~n}a, and Moss]{pasquazi2011sub}
Pasquazi,~A.; Peccianti,~M.; Park,~Y.; Little,~B.~E.; Chu,~S.~T.; Morandotti,~R.; Aza{\~n}a,~J.; Moss,~D.~J. Sub-picosecond phase-sensitive optical pulse characterization on a chip. \emph{Nature Photonics} \textbf{2011}, \emph{5}, 618--623\relax
\mciteBstWouldAddEndPuncttrue
\mciteSetBstMidEndSepPunct{\mcitedefaultmidpunct}
{\mcitedefaultendpunct}{\mcitedefaultseppunct}\relax
\EndOfBibitem
\bibitem[Riek \latin{et~al.}(2017)Riek, Sulzer, Seeger, Moskalenko, Burkard, Seletskiy, and Leitenstorfer]{riek2017subcycle}
Riek,~C.; Sulzer,~P.; Seeger,~M.; Moskalenko,~A.~S.; Burkard,~G.; Seletskiy,~D.~V.; Leitenstorfer,~A. Subcycle quantum electrodynamics. \emph{Nature} \textbf{2017}, \emph{541}, 376--379\relax
\mciteBstWouldAddEndPuncttrue
\mciteSetBstMidEndSepPunct{\mcitedefaultmidpunct}
{\mcitedefaultendpunct}{\mcitedefaultseppunct}\relax
\EndOfBibitem
\bibitem[Wegener(2006)]{wegener2006extreme}
Wegener,~M. \emph{Extreme nonlinear optics: an introduction}; Springer Science \& Business Media, 2006\relax
\mciteBstWouldAddEndPuncttrue
\mciteSetBstMidEndSepPunct{\mcitedefaultmidpunct}
{\mcitedefaultendpunct}{\mcitedefaultseppunct}\relax
\EndOfBibitem
\bibitem[Li \latin{et~al.}(2024)Li, Leefmans, Williams, Gray, Parto, and Marandi]{li2024deep}
Li,~G.~H.; Leefmans,~C.~R.; Williams,~J.; Gray,~R.~M.; Parto,~M.; Marandi,~A. Deep learning with photonic neural cellular automata. \emph{Light: Science \& Applications} \textbf{2024}, \emph{13}, 283\relax
\mciteBstWouldAddEndPuncttrue
\mciteSetBstMidEndSepPunct{\mcitedefaultmidpunct}
{\mcitedefaultendpunct}{\mcitedefaultseppunct}\relax
\EndOfBibitem
\bibitem[Trebino(2000)]{trebino2000frequency}
Trebino,~R. \emph{Frequency-Resolved Optical Gating: The Measurement of Ultrashort Laser Pulses: The Measurement of Ultrashort Laser Pulses}; Springer Science \& Business Media, 2000\relax
\mciteBstWouldAddEndPuncttrue
\mciteSetBstMidEndSepPunct{\mcitedefaultmidpunct}
{\mcitedefaultendpunct}{\mcitedefaultseppunct}\relax
\EndOfBibitem
\bibitem[Zhang \latin{et~al.}(2003)Zhang, Shreenath, Kimmel, Zeek, Trebino, and Link]{zhang2003measurement}
Zhang,~J.-y.; Shreenath,~A.~P.; Kimmel,~M.; Zeek,~E.; Trebino,~R.; Link,~S. Measurement of the intensity and phase of attojoule femtosecond light pulses using optical-parametric-amplification cross-correlation frequency-resolved optical gating. \emph{Optics Express} \textbf{2003}, \emph{11}, 601--609\relax
\mciteBstWouldAddEndPuncttrue
\mciteSetBstMidEndSepPunct{\mcitedefaultmidpunct}
{\mcitedefaultendpunct}{\mcitedefaultseppunct}\relax
\EndOfBibitem
\bibitem[Zhang \latin{et~al.}(2004)Zhang, Lee, Huang, and Pan]{zhang2004sub}
Zhang,~J.-Y.; Lee,~C.-K.; Huang,~J.~Y.; Pan,~C.-L. Sub femto-joule sensitive single-shot OPA-XFROG and its application in study of white-light supercontinuum generation. \emph{Optics Express} \textbf{2004}, \emph{12}, 574--581\relax
\mciteBstWouldAddEndPuncttrue
\mciteSetBstMidEndSepPunct{\mcitedefaultmidpunct}
{\mcitedefaultendpunct}{\mcitedefaultseppunct}\relax
\EndOfBibitem
\bibitem[Ledezma \latin{et~al.}(2022)Ledezma, Sekine, Guo, Nehra, Jahani, and Marandi]{ledezma2022intense}
Ledezma,~L.; Sekine,~R.; Guo,~Q.; Nehra,~R.; Jahani,~S.; Marandi,~A. Intense optical parametric amplification in dispersion-engineered nanophotonic lithium niobate waveguides. \emph{Optica} \textbf{2022}, \emph{9}, 303--308\relax
\mciteBstWouldAddEndPuncttrue
\mciteSetBstMidEndSepPunct{\mcitedefaultmidpunct}
{\mcitedefaultendpunct}{\mcitedefaultseppunct}\relax
\EndOfBibitem
\bibitem[Jankowski \latin{et~al.}(2022)Jankowski, Jornod, Langrock, Desiatov, Marandi, Lon{\v{c}}ar, and Fejer]{jankowski2022quasi}
Jankowski,~M.; Jornod,~N.; Langrock,~C.; Desiatov,~B.; Marandi,~A.; Lon{\v{c}}ar,~M.; Fejer,~M.~M. Quasi-static optical parametric amplification. \emph{Optica} \textbf{2022}, \emph{9}, 273--279\relax
\mciteBstWouldAddEndPuncttrue
\mciteSetBstMidEndSepPunct{\mcitedefaultmidpunct}
{\mcitedefaultendpunct}{\mcitedefaultseppunct}\relax
\EndOfBibitem
\end{mcitethebibliography}

\end{document}